\documentclass[journal]{vgtc}                

\ifpdf
  \pdfoutput=1\relax                   
  \usepackage{graphicx}                
  \DeclareGraphicsExtensions{.pdf,.png,.jpg,.jpeg} 
\else
  \usepackage{graphicx}                
  \DeclareGraphicsExtensions{.eps}     
\fi%

\graphicspath{{figures/}{pictures/}{images/}{./}} 

\usepackage{microtype}                 
\PassOptionsToPackage{warn}{textcomp}  
\usepackage{textcomp}                  
\usepackage{mathptmx}                  
\usepackage{times}                     
\usepackage{cite}                      
\usepackage{tabu}                      
\usepackage{booktabs}                  
\usepackage{paralist}
\usepackage{enumitem}

\usepackage{setspace}

\usepackage{multirow}
\usepackage{color}
\usepackage{amsmath}
\usepackage{bm}
\usepackage{xspace}

\onlineid{1104}

\vgtccategory{Research}
\vgtcpapertype{algorithm/technique}

\title{CausalFlow: Visual Analytics of Causality in Event Sequences}

\author{Xiao Xie, Moqi He, and Yingcai Wu}
\authorfooter{
\item
  X.Xie, M.He, and Y.Wu are with the State Key Lab of CAD\&CG, Zhejiang University. E-mail: \{xxie, moqihe, ycwu\}@zju.edu.cn
}

\shortauthortitle{Biv \MakeLowercase{\textit{et al.}}: Global Illumination for Fun and Profit}

\abstract{
Understanding the relation of events plays an important role in different domains, such as identifying the reasons for users' certain actions from application logs as well as explaining sports players' behaviors according to historical records. Co-occurrence has been widely used to characterize the relation of events. However, insights provided by the co-occurrence relation are vague, which leads to difficulties in addressing domain problems. In this paper, we use causation to model the relation of events and present a visualization approach for conducting the causation analysis of event sequences. We integrate automatic causal discovery methods into the approach and propose a model for detecting event causalities. Considering the interpretability, we design a novel visualization named causal flow to integrate the detected causality into timeline-based event sequence visualizations. With this design, users can understand the occurrence of certain events and identify the causal pathways. We further implement an interactive system to help users comprehensively analyze event sequences. Two case studies are provided to evaluate the usability of the approach.
} 


\CCScatlist{ 
 \CCScat{K.6.1}{Management of Computing and Information Systems}%
{Project and People Management}{Life Cycle};
 \CCScat{K.7.m}{The Computing Profession}{Miscellaneous}{Ethics}
}

\vgtcinsertpkg

\begin{document}



\maketitle

\begin{spacing}{0.97}

\section{Introduction}

The causation analysis of event sequence data can characterize the relation between events, thereby playing an important role in various domains, such as the behavior analysis of marketing\cite{kdd/AttenbergPS09} (e.g., factors that leading to users’ purchase), the electronic medical records (EMR) analysis of healthcare\cite{ihi/TaoWCPSC12} (e.g., reasons for patients having certain symptoms), and the error log analysis of industry\cite{infovis/ElmqvistT03} (e.g., reasons for system failure). Although controlled experiments are regarded as an appropriate method for deriving the causation of events, these experiments are not applicable in numerous scenarios owing to the high cost of experiment setting.

\begin{figure}[htb]
	\centering
	\includegraphics[width=1\linewidth]{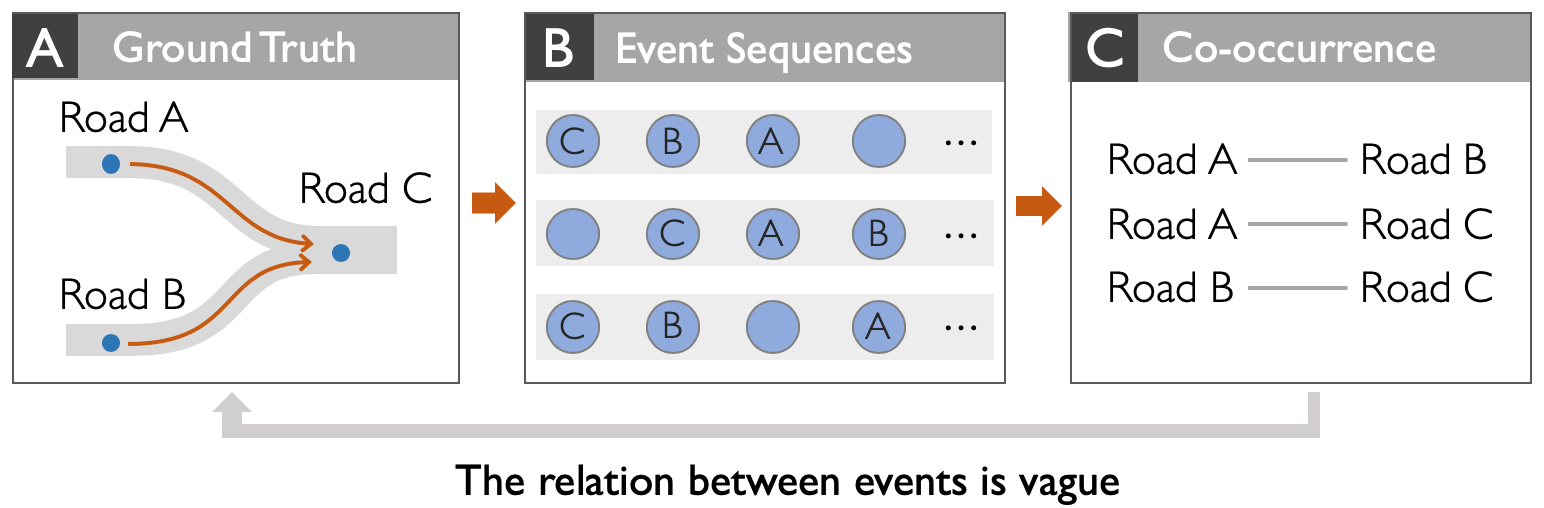}
    \caption{Co-occurrence relationships cannot fully describe the relation between events.}~\label{motivation}
    \vspace{-4mm}
\end{figure}

Due to the difficulties of identifying causation, most of prior studies \cite{outflow, careflow, decisionflow} use co-occurrence to model the relation between events.
Co-occurrence based approaches assume events that frequently co-occur in a sequence to be highly related.
However, co-occurrence is not causation. The insights provided by co-occurrence are comparatively vague.
For example, for the analysis of traffic congestion (Fig.~\ref{motivation}(A)), it is common to see that the congestion of a road (C) causes the congestion of other roads (A, B). Hence, it is critical for analysts to find the ``root congestion''.
Based on the event sequences of congestion (Fig.\ref{motivation}(B)), analysts can identify multiple congestion patterns with the co-occurrence approach (Fig.\ref{motivation}(C)).  
However, discovering the ``root congestion'' from these patterns is difficult since there is no direction for the co-occurrence relationship. 
Therefore, using causation to characterize the event relationship remains indispensable.

This study intends to develop a novel approach for conducting a causation analysis of event sequence data. This approach is two-fold. The first part is a causality detection module. We decide to use automatic causality detection methods to help users identify event causation. The second part is an interactive visual analytics system for conducting causation analysis. Given the causality detection module, the causation analysis continues to require human intelligence for validating and interpreting discovered event causations.

We face three challenges when developing this approach. The first challenge is the causal detection of event sequence data. The existing causality detection methods are mainly designed for tabular data and cannot be directly applied to event sequences. Compared with tabular data, event sequences are time-variant, and the causation of events could be changed over time. A typical example is the relation between users’ online behaviors and purchases. Users’ decision-making processes related to purchases vary according to their different personal states. Accordingly, how to handle the time-variant causation of events remains an important issue.

The second challenge is the problem characterization of the causation analysis of event sequences. Although previous visualization research has involved the different aspects of analyzing event causation, a thorough study of conducting causation analysis of event sequences remains lagging. Moreover, automatic causality detection methods were integrated for the first time into the visual analytics system for event sequence analysis. Hence, a comprehensive study of the problem domain should be conducted. The third challenge is integrating causality information into event sequence visualizations. Event sequences are often visualized using a  timeline-based visual representation. However, showing the causality information of events to explain the occurrence while preserving the chronological order of events in visualizations has yet to be fully explored.

To address the first challenge, we propose a causal detection framework for event sequence data. We use an iterative framework to extract multiple causal graphs for addressing the time-variant issue and propose a data transformation method for applying causal detection techniques to event sequence data. To address the second challenge, we propose a novel visual design that combines Sankey diagrams and timeline-based visualizations to present the causality of event sequences.
A force-directed-based layout approach is provided to reduce visual clutter and ensure the readability of different causal structures.
To address the third challenge, we discuss with experts from different domains and distill 6 design requirements. We further design and implement a visualization system according to the requirements. 

The contributions of this study are as follows:
\begin{itemize}
    \item Efficient algorithms to identify and model the multiple causal states within event sequences.
    \item Generic visual summary of event sequences that combines the frequent sequential patterns and the causality information between events.
    \item Interactive visualization system that can support users to fully explain the occurrence of an event with event sequence. 
\end{itemize}
\section{Related Work} \label{sec:related_work}
This work is mainly related to the visual analytics of causality, the detection of causality, and the analysis of event sequences.

\subsection{Visual Analytics of Causality}
Early studies on causality visualizations have focused on using visualizations to help users recognize the causality information.
One representative is the Hasse diagram, which has been frequently used to present the interactions between events in a parallel process or a distributed system. However, this approach is not scalable when the distributed system is large. Therefore, Growing squares\cite{ivs/ElmqvistT04} and Growing polygons\cite{infovis/ElmqvistT03}, which use animation techniques rather than a linear timeline, are proposed to address the scalability issue. ReactionFlow\cite{reactionflow} provides an interactive tool for analyzing the domain-specific causality of biochemical interactions. Downstream and upstream relations are visualized to help users identify meaningful causal pathways.

In recent years, by introducing the theory of representing causality as a Bayesian network\cite{ijon/Shanmugam01}, researchers have turned to study suitable graph visualization techniques for visualizing the causality that already known. Kadaba et.al.\cite{tvcg/KadabaIL07} create different glyphs and place them on the directed acyclic graph (DAG) to show the different attributes (e.g., the strength) of causality. A similar work that focused on the causation analysis of events is proposed by Vigueras and Botia \cite{vigueras2007tracking}. They use causal graphs to visualize the causation of events in a multi-agent system. Yet the causation is defined by the order of events which is domain-specific and cannot be applied to other domains. Bae et al. \cite{cgf/BaeHR17} focus on the problem of distinguishing between direct and indirect causal relations, and conduct an experiment to study the effectiveness of a sequential graph layout for showing causalities. For these studies, causality has often been assumed provided or defined according to domain knowledge. Hence, the preceding studies cannot be directly applied to the causation analysis of event sequences.

As the first work that integrates automatic causality detection techniques with visualization, Wang et.al.\cite{tvcg/WangM16} use a force-directed graph layout to visualize the detected causal network and develop an interactive system for users to explore and adjust the causal network on demand. They further provide a flow-based ƒ graph visualization for recognizing the causal pathways \cite{tvcg/WangSAZZYQ16}. However, these works focus on visualizing the causality discovered from static multi-dimensional data. None of existing causal visualizations have explored the visual representation for integrating detected causalities with event sequence visualizations for explaining the occurrence of events. In this paper, we intend to fill this gap by designing causation-aware event sequence visualizations.

\subsection{Causality Detection}
Causal detection techniques can be divided into two categories according to the type of input data, i.e., one for static variables and the other for temporal variables. For the static variables, numbers of methods have been proposed based on Pearl's structural causal model. These methods aim to find a Bayesian network (causal network) from a multi-dimensional dataset. In the network, nodes encode the dimension and the directed edges encode the direction of causation between dimensions. Methods like SGS\cite{sgs/spirtes1991}, PC\cite{pc/spirtes1991, pc/colombo2014order, pc/spirtes2000causation}, and TC\cite{tc/pellet2008using} discover the causal network by testing the conditional independence (CI tests) between dimensions. GES\cite{jmlr/Chickering02a} use a greedy search method to address the network search problem and propose a score function as a likelihood estimation of the causal network. Goudet et.al.\cite{Goudet2018} further propose a differentiable function for causality detection and use a deep learning model to address this problem. 

For the temporal variable (i.e., a time series, such as the condition of weather recorded each day), finding the causality is to validate whether the value of time series $X = [x_0, x_1, ..., x_t]$ is helpful for forecasting the value of time series $Y = [y_0, y_1, ..., y_t]$. Granger causality test\cite{granger} is a popular method for detecting such temporal causalities and has been widely adopted in different domains.

However, event sequences are regarded as a complex dataset that cannot be easily classified as static or temporal variables. Compared with static variables, each event contains additional temporal information, which cannot be neglected while compared with temporal variables. The records of different events also cannot be aligned as the events happen at different timestamps. Hence, aforementioned methods cannot be directly applied to detect the causality of event sequences.
 
 \subsection{Event Sequence Analysis}
 Over the past years, visualizing event sequences has received substantial attention and applied to different domains\cite{matrixwave, peerfinder, chi/GuoDMKKLKZC19, eventaction}. Early works like Lifelines\cite{lifelines} and Cloudlines\cite{cloudlines} visualize raw event sequences with a timeline-based visual representation. Allowing users to conduct a detailed analysis of multiple individual event sequences, these visualizations face scalability issues when the number of event sequences grows. Therefore, a set of methods have been proposed to provide an overview for large-scale event sequences. EventFlow\cite{eventflow1, eventflow2} and LifeFlow\cite{lifeflow} organize events in a hierarchical structure and visualize them with Icicles. Given the problem of noisy data in event sequences and the redundant sequential patterns, Chen et.al.\cite{tvcg/ChenXR18} propose a visual summarization technique that considers both the visual simplicity and the information loss based on the information theory. CoreFlow\cite{coreflow} proposes methods for detecting branching patterns in event sequences and create a tree-based representation for visualizing such patterns. Focusing the issue of identifying event pathways, OutFlow\cite{outflow} and CareFlow\cite{careflow} use a sankey diagram to visualize the transition between events. Different event pathways that result in an eventual outcome are presented to help users figure out how and why certain event sequences could lead to a specific outcome. DecisionFlow\cite{decisionflow} further provides a scalable approach to visualize pathways of event sequences when there are a large number of events.

\begin{figure*}[htb]
	\centering
	\includegraphics[width=1\linewidth]{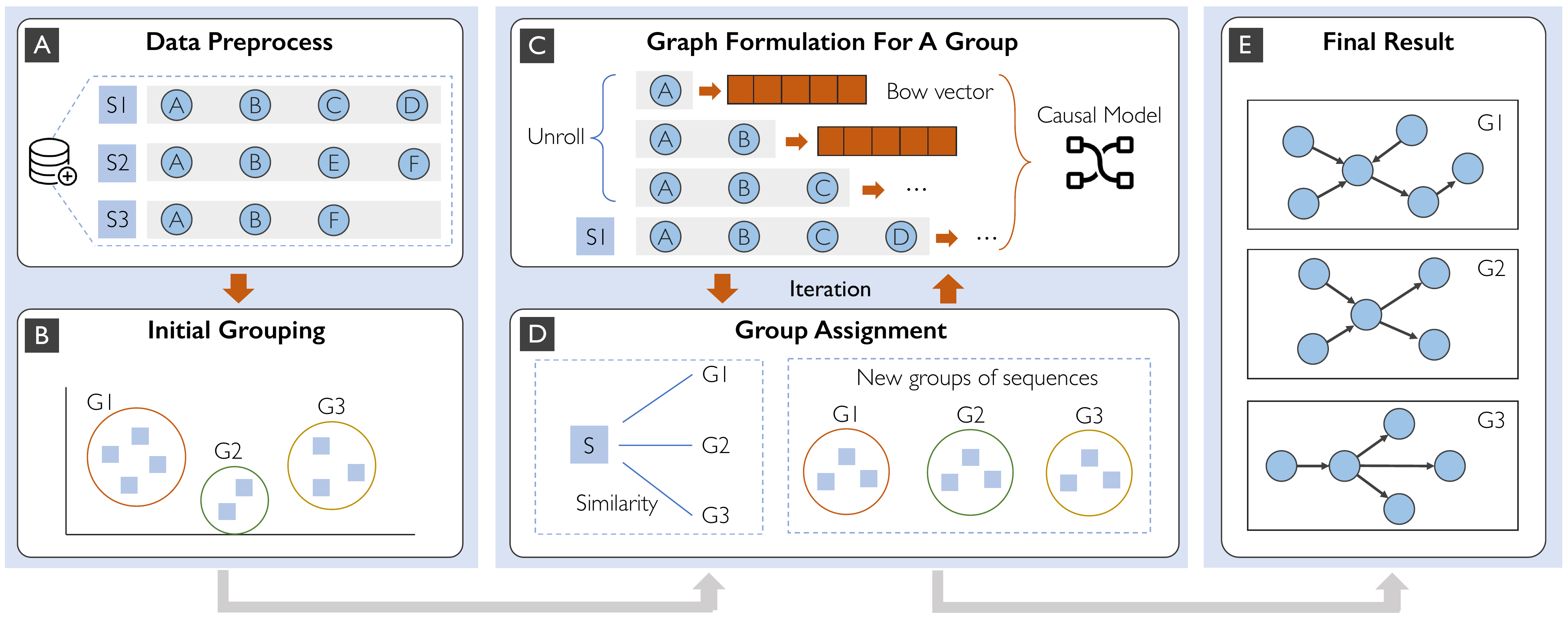}
    \caption{The pipeline of the causal detection model. (A) Obtain event sequences from raw data. (B) Build initial groups of event sequences by clustering. (C) Detect the causal graphs of each group of event sequences. (D) Assign each event sequence to a causal graph that can best explain the sequence. (E) The final causal graphs of each group after the convergence of the iterative process between (C) and (D).}~\label{model}
    \vspace{-4mm}
\end{figure*}

Compared with previous works, a significant difference is that we use automatic causal detection techniques to detect the causality and use visualizations to directly present how the causality can produce the observed event sequences.

\section{Modeling Event Causality} \label{sec:model}
In this section, we first give a brief description of the data and the theory of causal networks.
Thereafter, we demonstrate problems of detecting the causality in event sequences. Lastly, we present a framework for addressing the problems.

\subsection{Data description}
An event sequence is an ordered sequences of events that can be frequently seen in different domains. For example, users' interactions on a software will be recorded as event sequences. For a formal description, we have an event set $D = \{d_1, d_2,..., d_m\}$ that contains the $m$ possible events in event sequences and a sequence set $S = \{S_1, S_2, ..., S_n\}$ with $n$ event sequences, $S_i = [ e_1^d, e_2^d, ..., e_l^d ]$. Before detecting the causality of events, we filter out noisy events and merge successive identical events as one event (e.g., $[e_i^j, e_{i+1}^j] = [e_i^j]$) to reduce the complexity.

\subsection{Causality Representation}
\label{causal_detection}
The most popular causality representation introduced by Pearl's theory\cite{ijon/Shanmugam01} is to encode causalities as a Bayesian network $G = (V, E)$. In this network, each node represents a dimension and the directed edge between nodes encodes the causal direction. For example, node $A$ has an edge pointing to node $B$ only when $A$ is the cause of $B$. From a statistical perspective, an edge between $A$ and $B$ means that $A$ is not independent of $B$ under any controlled conditions, i.e.,
\begin{equation}
\label{eq1}
P(AB|Z) \neq P(A|Z) * P(B|Z), \forall Z \subseteq  V_{\backslash \{A, B\}}
\end{equation}
For any two nodes without a connecting edge, the statistical situation is 
\begin{equation}
\label{eq2}
P(AB|Z) = P(A|Z) * P(B|Z), \exists Z \subseteq  V_{\backslash \{A, B\}}
\end{equation}
where $V_{\backslash \{A, B\}}$ is the set of all dimensions except $A$ and $B$.  According to (\ref{eq1}) and (\ref{eq2}), the joint distribution $P$ over all the dimensions can be factorized based on the structure of the causal graph:
\begin{equation}
P = \prod_{i=1}^{n}P(X_i|Parent(X_i))
\end{equation}
where $n$ is the total number of nodes, $X_i$ is a node, and $Parent(X_i)$ is the parents of $X_i$. A set of causal discovery techniques have been proposed based on these characteristics. Here we will briefly introduce the classic causal discovery algorithm PC (Peter - Clark)\cite{pc/spirtes1991}.

The overall idea of PC\cite{pc/spirtes1991} is to delete edges that cannot be a causal relation. Given a tabular dataset with $r$ rows and $n$ dimensions, PC will create a fully connected graph with $n$ nodes as the initial causal graph. Based on the characteristics of (\ref{eq1}) and (\ref{eq2}), PC will reduce the edges by testing the conditional independence between every two nodes. If for $A$ and $B$, a condition set $Z$ is found such that (\ref{eq2}) is satisfied, the edge between $A$ and $B$ will be deleted from the causal graph. In practice, the conditional independence can be computed by the partial correlation\cite{ci/baba2004partial}. Partial correlation is used to estimate the correlation of two variables while controlling for the effect of a controlled set of variables. Therefore, partial correlation is regarded as an accurate estimation of the directed effect (i.e., the effect that variable A is affected directly by B with no mediation) between variables. Given a set $Z$ and a correlation matrix $\Omega$ of the variables, the partial correlation between $X_i$ and $X_j$ conditioning on $Z$ is defined as:
\begin{equation}
\rho_{X_iX_j{\cdot}Z} = -\frac{p_{ij}}{\sqrt{p_{ii}p_{jj}}}
\end{equation}
where $p_{ij}$ is from the precision matrix $P = {\Omega}^{-1}$. PC has provided methods for reducing the numbers of conditional independence tests to accelerate the causal discovery process. Through this manner, a set of edges is preserved which are regarded as the possible causal relations. The final step is to orient the causal direction of each edge. Although this step is important for the causal discovery, this aspect is not the focus of this paper. Additional details of the orientation can be found in \cite{pc/spirtes1991}. Noted that not all the edges can be assigned an orientation. We only use the edges that the algorithm is sure about the orientation.

\subsection{A Framework for Discovering Event Causality}
\begin{figure}[htb]
	\centering
	\includegraphics[width=1\linewidth]{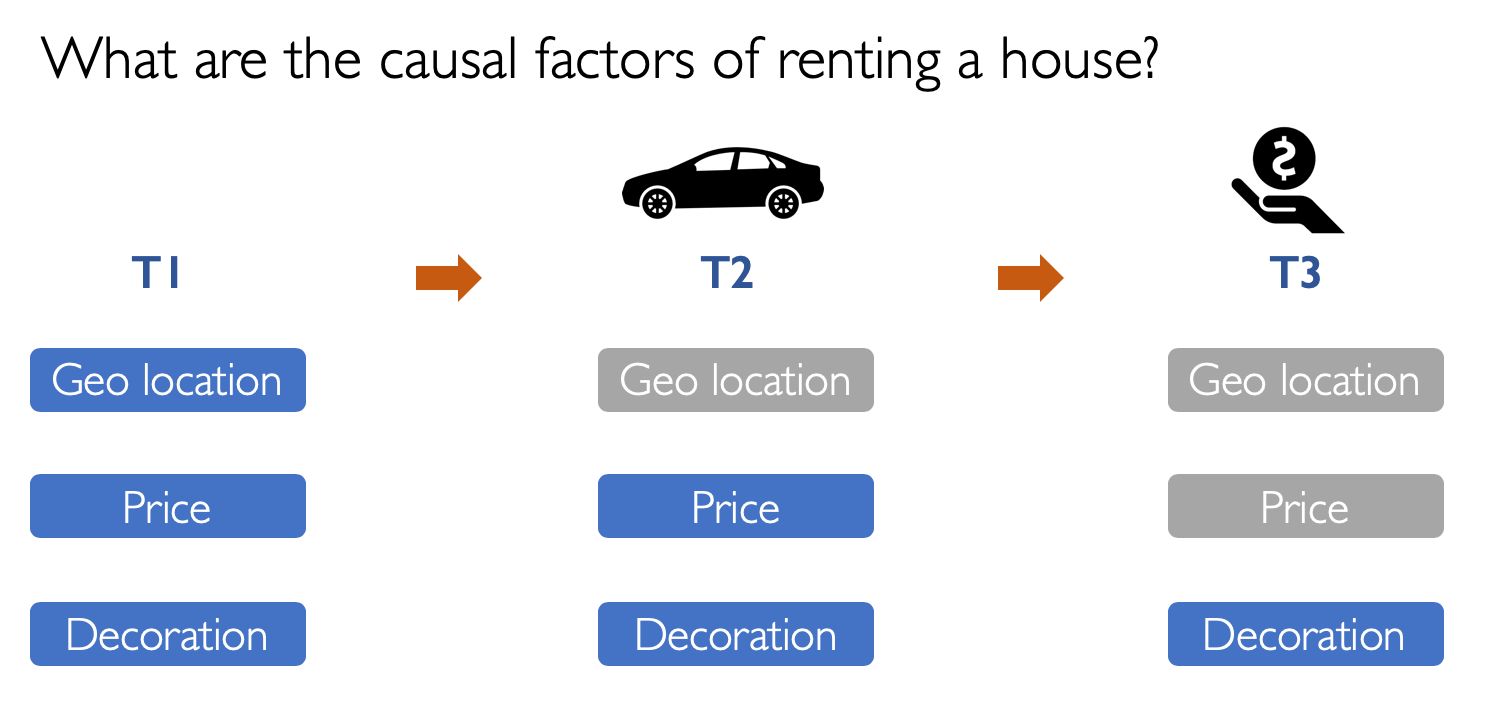}
    \caption{The possible causal states in event sequences. At T1, T2, and T3, factors that majorly influence Bob's purchase decision are different due to the changing environment. Therefore, the causality of event sequences cannot be represented by one causal graph.}~\label{causal_state}
    \vspace{-4mm}
\end{figure}

Discovering the event causality is to find an appropriate causal graph of events that can be used to explain the generation of certain event sequences. However, using only one causal graph to explain the generation of a long event sequence is difficult because the relation of events would change over time. Suppose we have a customer Bob who is preparing house renting. At $T1$ (Fig.~\ref{causal_state}), Bob would like to find a house that is close to his working position with affordable prices and tolerable decoration. Hence, the main factors that affect his decisions are the geo-location, the price, and the decoration. However, as time goes by, at $T2$, the main factors that affect A's purchase decisions may be the price and decoration. The reason is that at $T2$ (Fig.~\ref{causal_state}) Bob has bought a car and the geo-location now is not a main constraint for selecting a house. The main factors of renting a house will further change according to A's dynamic personal states at $T3$ and $T4$ (Fig.\ref{causal_state}). Therefore, the causality related to the purchase is dynamic over time, thereby requiring multiple causal graphs to explain the user’s action sequences.

\textbf{Problem Definition.} According to the aforementioned scenario, an event sequence $S_i$ can be divided into sessions $[{\S}_1^i, {\S}_2^i, ..., {\S}_n^i]$. In practice, we derive these sessions by setting time intervals. Each session ${\S}_j^i$ is a sequence of events whose generation can be explained by a causal graph $G_i$. Therefore, we regard the causality detection of event sequences as a grouping problem. Given a dataset $S$, the target is to identify a set of causal graphs $\{ G_i\}$ that presents the event causality and a mapping function $f$ that assigns each ${\S}_j^i$ to a causal graph $G_i$. 

Inspired by traditional clustering methods, we use a kmeans-like approach to address this problem. 
As shown in Fig.~\ref{model}, we first derive event sequences data (Fig.~\ref{model}(A)) and use clustering to group the event sequences (Fig.~\ref{model}(B)). 
We transform each sequence to a vectorized embedding and perform a density-based clustering based on the embedding to find an appropriate $K$ as the number of causal graphs.
This is inspired by Wang et.al.\cite{ieeevast/WangM17} who use clustering to find the different causal groups in a dataset.
Later, the framework will iteratively run the \textbf{Formulating} step (Fig.~\ref{model}(C)) and the \textbf{Grouping} (Fig.~\ref{model}(D)) step to detect the event causality. 
The \textit{\textbf{Formulating}} step is designed to compute a causal graph $G$ for each group of sequences and the \textit{\textbf{Grouping}} step is to determine the group of a sequence given a groups of causal graphs $\{ G_i\}$.
The iterating process will stop when the causal groups are converged, i.e., the sequences of each group do not change after an iteration. 
Finally, we obtain a set of causal graphs that describe the potential causality of events (Fig.~\ref{model}(E)). 
The details of each step are provided as follows.

\textbf{Grouping.} Given a set of causal graphs $\{ G_i\}$, this step is to assign each ${\S}_j^i$ to a causal graph. This is similar to the assignment step of k-means that assigns each data point to the nearest cluster. Hence, a distance metric $d(G_i, {\S}_j^i)$ is required to find the nearest causal graph for each session. Here we define the distance as
\begin{equation}
d(G_i, {\S}_j^i) = \frac{1}{P({\S}_j^i|G_i)}
\end{equation}
where $P({\S}_j^i|G_i)$ is the probability of generating ${\S}_j^i$ given $G_i$. Hence, ${\S}_j^i$ will be assigned to the causal graph with the shortest distance.

\textbf{Formulating.} Given a set of sessions $\{{\S}_j^i\}$, this step is to identify a causal graph $G$ from the event records in these sessions. Since the given sessions are supposed to share the same causal relations, this step is analog to the regular causality detection process except that the input data is event records. As described in Sec.~\ref{causal_detection}, regular causality detection techniques, such as PC\cite{pc/spirtes1991}, TC\cite{tc/pellet2008using}, and GES\cite{jmlr/Chickering02a}, are usually used to identify the causal network from a multi-dimensional tabular data. Although event records can be easily transformed into a tabular dataset to apply current causal detection techniques, the temporal information, i.e., the order of events, cannot be preserved. To address this issue, the framework processes the session data as shown in Fig.~\ref{model} (C). For an $n$ length event sequence ${\S}_j^i = [e_1, ..., e_n]$, the framework will transform it to $n$ subsequences $[s_1, ..., s_n]$ where $s_i = [e_1, ..., e_i]$. For each subsequence $s_i$, a feature vector is extracted correspondingly. Each feature vector is a bag-of-word representation of events. A tabular dataset is created based on the feature vectors from all sessions. The causal graph $G$ can be obtained by applying causal discovery algorithms on this tabular dataset.

\section{Background} \label{sec:interview}

In this section, we first describe how we collect and summarize tasks based on previous works and conduct interviews with different domain experts.
We then provide a problem characterization of the causal analysis of event sequence data.

\subsection{Interviews}

Understanding the generation of certain event sequences and the reason of the happening of specific events are common tasks shared among various domains. For example, in the healthcare domain, doctors would be interested to identify the reasons why patients would have certain symptoms given a sequence of medical records. In sports, coaches would like to know why players would have certain action sequence patterns to establish substantially useful tactics. Such a type of reasoning activities can be regarded as analyzing the causality within event sequences. To investigate this topic, we collected and summarized relevant studies in the area of visualizations and causal detection. We use the literature review as a basis to summarize a set of requirements for the causality analysis of event sequences. Moreover, we conducted brainstorming sessions and interviewed 10 users (3 females) to confirm the validity of the proposed requirements. Each interview requires approximately 30 min. The final requirements are as follows.

\subsection{Requirements}
\label{requirement}
Inspired by previous works and experts' feedback, we derive the following requirements for analyzing the causality of event sequences.

\begin{compactenum}[R1]
\item \textbf{How many causal groups does a dataset contain? How much data does each group cover?} Grouping event sequences according to the causal relations can help users obtain an overview of the data, which is useful when analyzing a large scale of event sequences. Moreover, knowing how many sequences each group cover can help users focus on the major causal group in a dataset.

\item \textbf{What are the causal relation of events for each causal group?} For a detailed analysis, users need to know the relation between events, i.e., what are the causes of each event and how these events formulate a causal network. Users can obtain knowledge and insights from the detected causal relations and can also validate the causal relations according to their domain experiences.

\item \textbf{How are the event sequences progress according to the causal relations?} It may be direct for users to understand the meaning of event causalities with the causal network alone. However, how to use the causal network to explain the progress of the observational event sequences data is still indirect. Hence, this is an important requirement for users to apply the detected causalities for understanding why certain event sequences would happen.

\item \textbf{Do different causal groups share similar event sequences?} As an event sequence can be generated by different causal states, for completeness, it is crucial for users to find out all the potential reasons for certain event sequences. This can help users reduce the bias understanding and find out the most possible reason, which is helpful for users' decision-making process.

\item \textbf{What are the possible causal pathways of an event?} It is common to see that users want to reduce or increase the number of certain events when conducting event sequence analysis. Hence, identifying and showing the pathways of an event based on the detected causal states can provide insights for such purposes.

\item \textbf{What are the major causal pathways of an event?} When there are many different pathways that may lead to an event, users want to know the importance of different pathways to reduce the complexity of analysis. This can help users focus on the frequent pathways and provide suggestions for conducting interventions.
        
\end{compactenum}

\subsection{System overview}

\begin{figure}[h]
	\centering
	\includegraphics[width=1\linewidth]{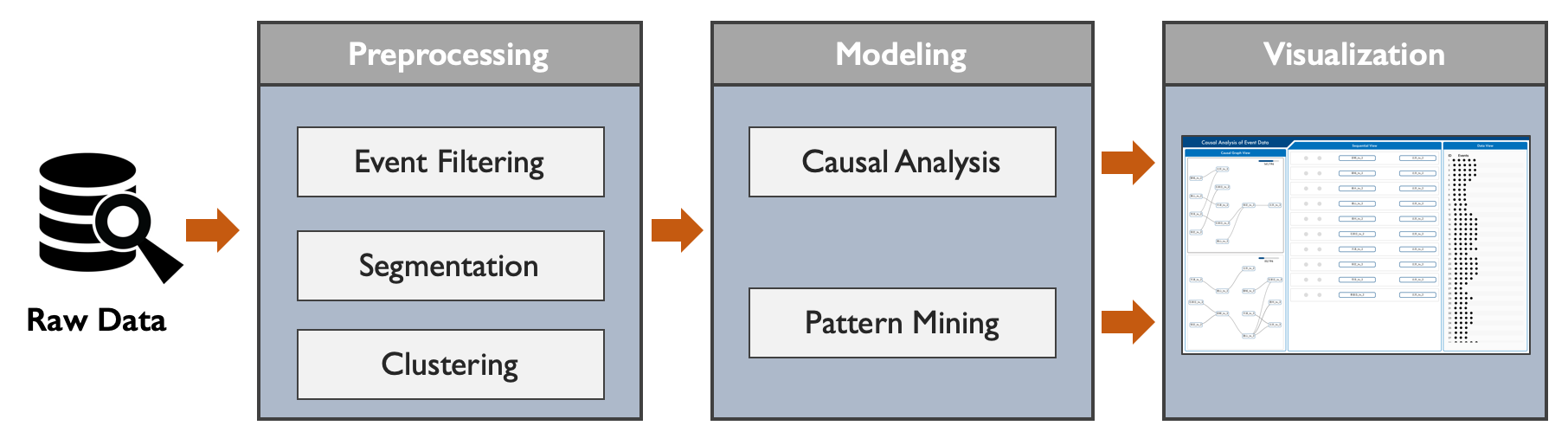}
    \caption{The system overview.}~\label{overview}
    \vspace{-4mm}
\end{figure}

As shown in Fig.~\ref{overview}, the system comprises of three components, a pre-processing component for handling the event sequence data, a modeling component for detecting the event causality and the sequential patterns, and a visualization component for conducting causation analysis. We use Python to implement the pre-processing and modeling component and adopt a causal discovery toolbox \cite{cdt} for the causal detection. We use Vue 2.0 to build the visualization component.

The visualization contains three views, namely, a causal graph view that presents the causal relations within event sequences, a sequential view that demonstrates how to use a causal graph to explain the generation of certain event sequential patterns, and a data view that help users inspect the raw event sequence data for the validation. 

\section{System} \label{sec:system}
In this section, we first present the design goals of our system and describe how we implement the system based on the goals.

\begin{figure*}[!htb]
	\centering
	\includegraphics[width=1\linewidth]{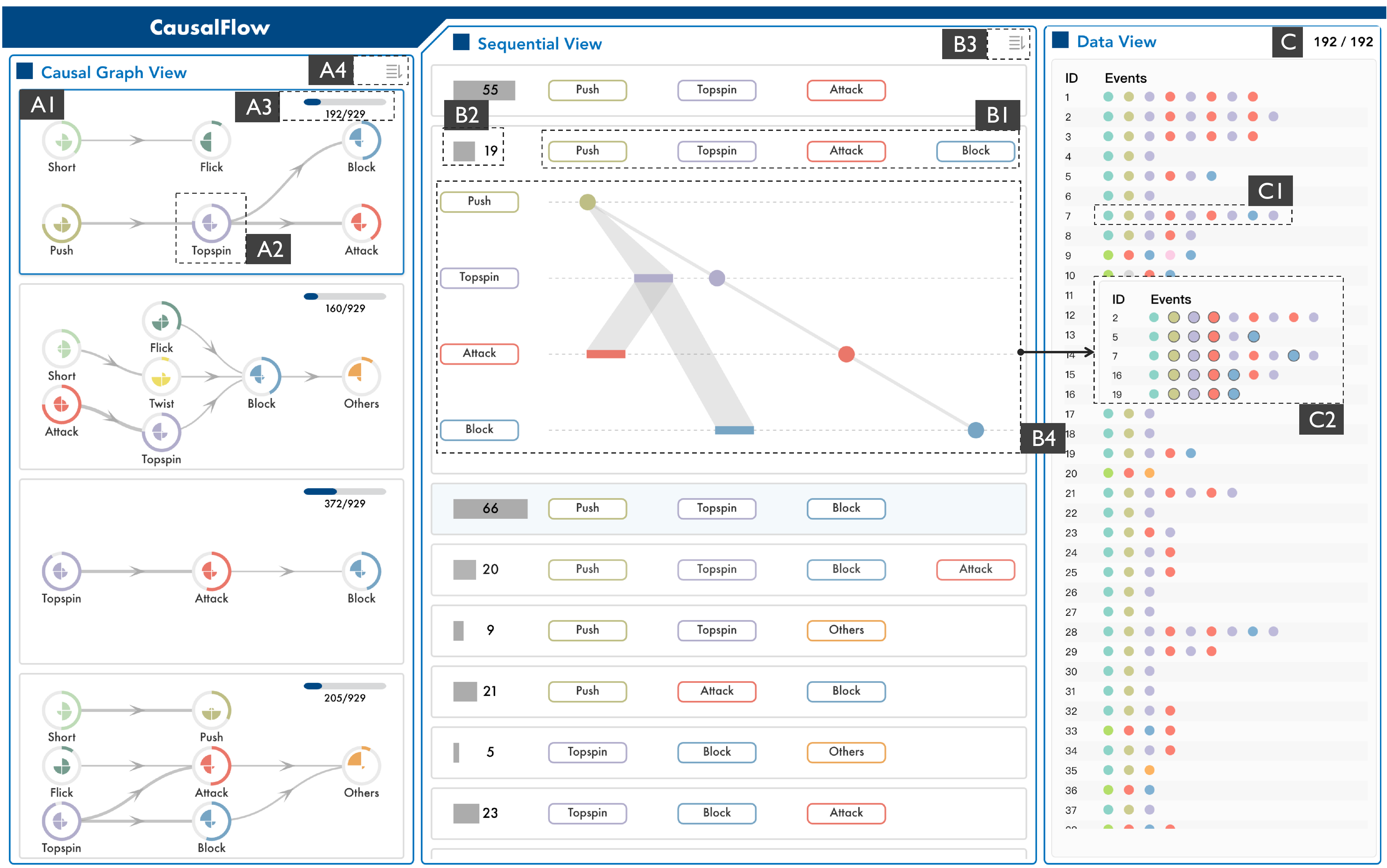}
    \caption{The system interface. (A) The causal graph view for exploring and comparing the causal graphs (A1). (B) The sequential view for inspecting and explaining the event sequential patterns (B1) of a selected causal graph. (C) The data view for validating the raw event sequences (C1).}~\label{system}
    \vspace{-4mm}
\end{figure*}

\subsection{Design Goals}
Based on the requirements, we derived the following design goals.
\begin{compactenum}[G1]
\item \textbf{Multi-level exploration of event sequences.} This is proposed for R1-R6. Users usually conduct causal analysis with a large scale of event sequences. More importantly, during the analysis, users would need to involve different concepts, i.e., causal graphs, sequential patterns, and raw sequences, which is a complex scenario. Hence, a multi-level exploration can help users better organize their analysis and significantly increase the efficiency. 
\item \textbf{Visual comparison of event causality.} In addition to showing the causality of each causal graph, providing visual comparison between graphs is important as this can help users identify common causal relations that shared by different people as well as exclusive causal relations of a specific group. Users can therefore obtain new insights and knowledge based on the comparison. (R1, R2)
\item \textbf{Timeline-based visual representation with causal information.} Timeline-based visual representations have been frequently used and widely accepted when visualizing event sequence data. Hence, this is the preferred layout for most users. Moreover, the causal information of events should be integrated into the timeline-based sequence visualization rather than presented separately. This can help users better understand the meaning of causality of events as well as the generation of event sequences. (R3, R4)
\item \textbf{Flow-based visualization of targeted event.} As there may be multiple pathways to a targeted event, a flow-based visualization can help users organize the different pathways in a view seamlessly, which is useful for a quick summarization. Moreover, the split and merge of flows can help users identify critical events in the pathways. It is also helpful when users need to compare the characteristics of different pathways. (R5, R6)
\end{compactenum}

\subsection{Visual Design}

The system is designed to support a multi-level exploration (G1). When analyzing event sequences with this system, users will first explore the causal graph view (Fig.~\ref{system}(A)).  The causal graph view shows multiple detected causal graphs and users can inspect the characteristics of each graph (G2, G4). By finding causal graphs of interest, users can further turn to the sequential view (Fig.~\ref{system}(B)) to see the sequential patterns created by the selected causal graph. A novel design called causal sequential flow is used to explain how the causality of events leads to the generation of certain sequential patterns (G3). For further validation, users can select a specific sequential pattern to see the raw event sequences that contain the pattern in the data view (Fig.~\ref{system}(C)). Details of each view are as follows.

\subsubsection{Causal Graph View}

This view shows the detected causal graphs of corresponding causal groups. As shown in Fig.~\ref{system}(A1), we follow the layout algorithm proposed by Wang et.al.\cite{ieeevast/WangM17} and use a flow metaphor to visualize the causality between events as a path diagram (G2, G4). In the graph, a glyph represents an event and the position of each glyph is placed from left to right according to the order of causality. For example, the link pointed from \textit{Push} to \textit{Topspin} (Fig.~\ref{system}(A1)) means that the occurrence of \textit{Push} is a possible reason for the occurrence of \textit{Topspin}. A thicker link encodes a stronger causal relationship. There are multiple graph layout algorithms for visualizing the causal graph, such as the force-directed layout and the circular layout. Considering the importance of preserving the local structure of causal graphs (i.e., the causal link and its direction) rather than the overall topological information (e.g., communities of nodes), we decided to use this path diagram to visualize the causal graphs. 

\begin{figure}[!htb]
	\centering
	\includegraphics[width=1\linewidth]{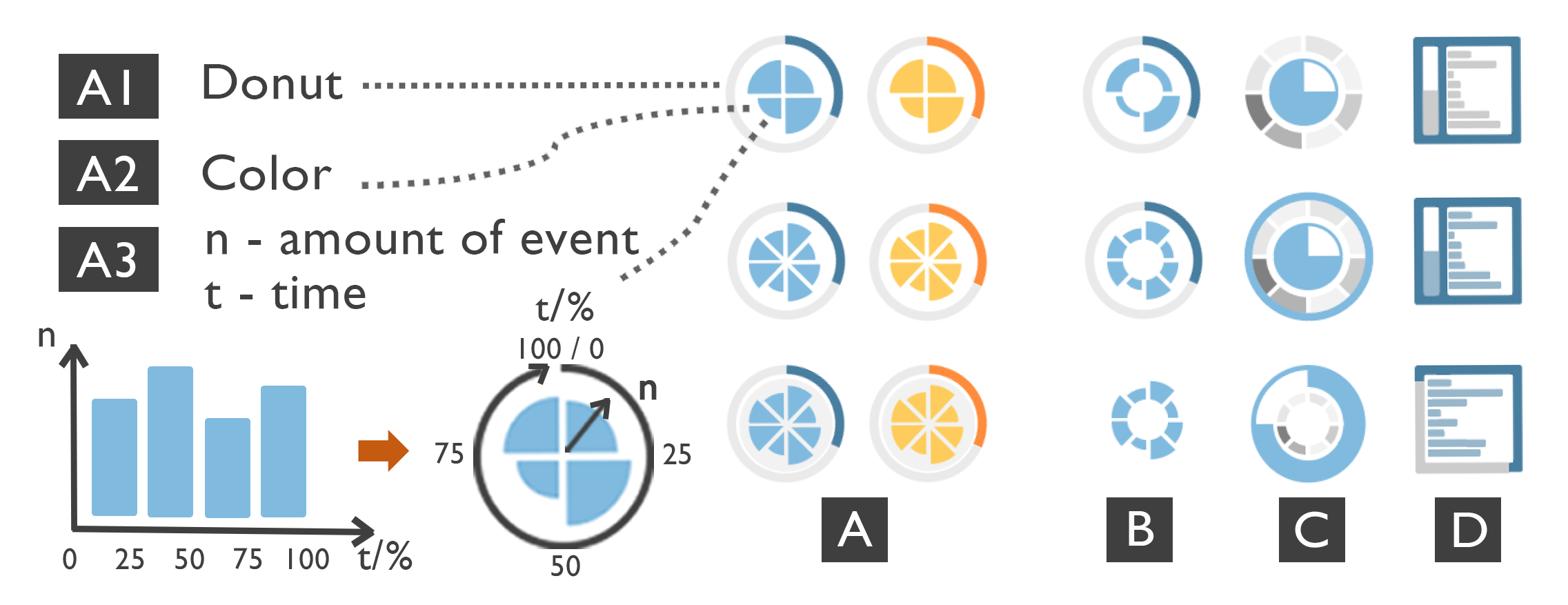}
    \caption{The encodings and alternatives of the event glyph. (A) is the glyph we current use. (B), (C), and (D) are the alternatives.}~\label{glyph}
    \vspace{-4mm}
\end{figure}

\textit{Event glyph.} As shown in Fig.~\ref{system}(A2), each event is represented by a glyph. The glyph is designed to present the multi-dimensional information of events. The design is guided by Maguire et.al's\cite{glyph} rationales. Specifically, we use the glyph to present three dimensions that are important for various domains, i.e., the event frequency, the event type, and the temporal distribution. The event frequency is encoded by the proportion of the donut (Fig.~\ref{glyph}(A1)). This can help users identify frequent events and rare events in the causal graph. 
The event type is encoded by the color (Fig.~\ref{glyph}(A2)). 
For example, we can use the color to distinguish between the different action events of players when analyzing events sequences of sports.
The temporal distribution is encoded by the size of the sector (Fig.~\ref{glyph}(A3)). 
We divide a sequence into quarters (Fig.~\ref{glyph}(A4)) and compute the distribution of events over quarters.
Therefore, users can obtain a summarization of the chronological information of events.
Users can hover on the glyph to see the detailed information.
We also design a set of alternatives for the event glyphs (Fig.~\ref{glyph}(B), (C), (D)). 
During the design process, we presented these glyphs to our users.
According to their comments, they prefer a circular layout (Fig.~\ref{glyph}(A), (B), (C)) as this is more intuitive for encoding a node.
Moreover, considering the efficiency of encoding proportion with donut charts and the large visual space for the temporal information, we decide to use Fig.~\ref{glyph}(A) as the design of the event glyph.

\begin{figure}[htb]
	\centering
	\includegraphics[width=1\linewidth]{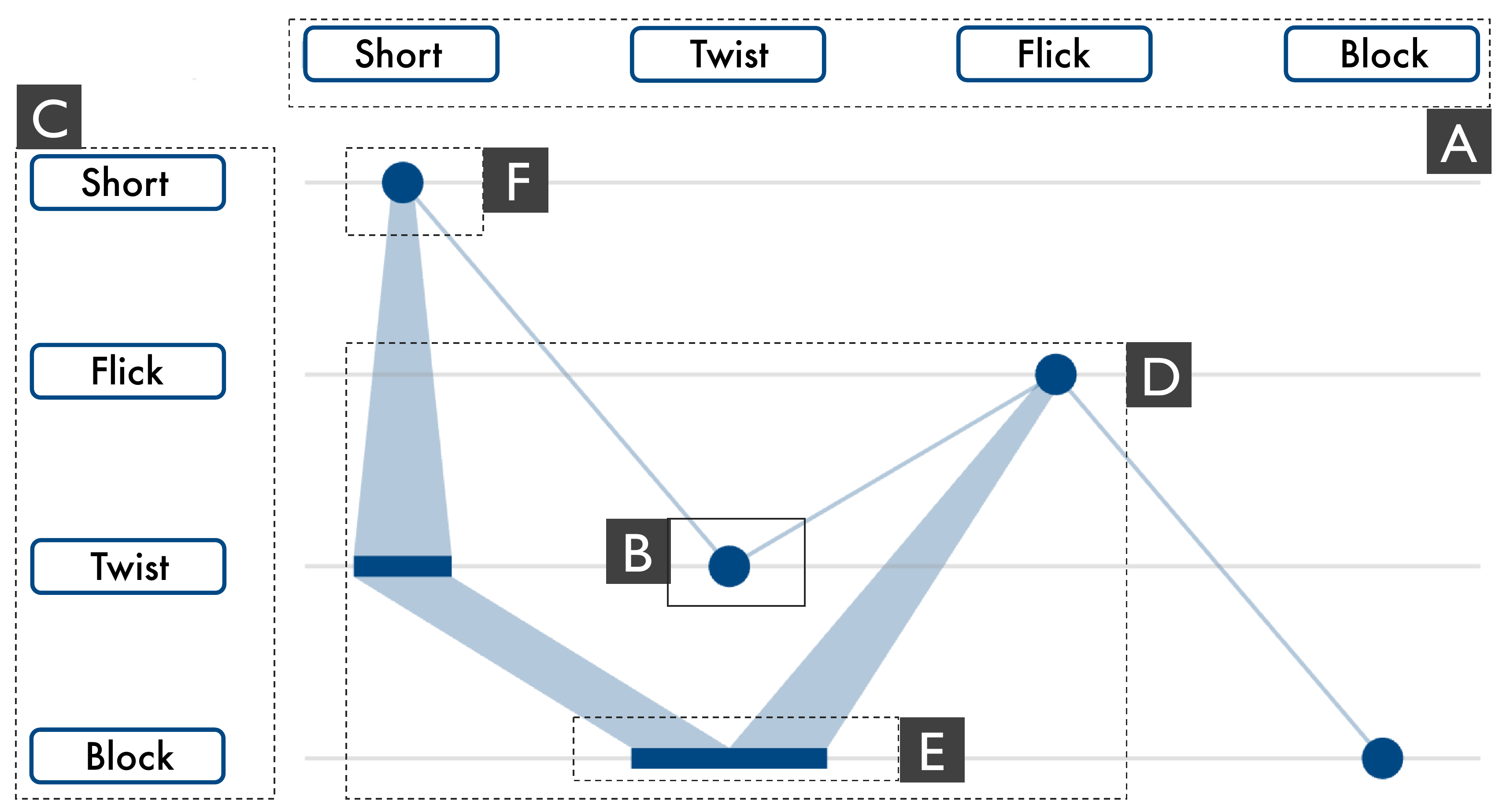}
    \caption{The design of causal sequential flow. (A) The sequential pattern of events. (B) The causal order of events based on the causal graph. (C) The event glyph corresponding to event C. (D) The sankey representation of showing the causes of event.}~\label{flow_design}
    \vspace{-4mm}
\end{figure}

Multiple causal graphs are juxtaposed in this view. To support the comparison between causal graphs (G2), we place a bar (Fig.~\ref{system}(A2)) on the right top of each causal graph to show the corresponding statistics information. The length of the bar encodes the number of sequences in each corresponding causal group, which can be regarded as the importance of each causal graph. A causal graph with a long bar means that this graph can explain most of the sequences in the dataset. Users can therefore sort (Fig.~\ref{system}(A3)) the causal graphs according to the bar. We further design interactions to support a detailed exploration and comparison. 
To clearly investigate a specific causal link, users can hover on the link and all the causal graphs that contain this link will be highlighted. Users can learn whether a causal link is suitable for a wide range or a small set of users. 

\subsubsection{Sequential View}
After exploring the causal graph view, users may be interested in a specific causal graph and would like to investigate how to use this causal graph to explain the observed sequences. To this end, users can click on a causal graph (Fig.~\ref{system}(A1)) to jump to the sequential view. As shown in Fig.~\ref{system}(B), in this view, users can first inspect the frequent sequential patterns (Fig.~\ref{system}(B1)) detected from the covered sequences of the selected causal graph. Each rect represents an event and the chronological order of events is from left to right. The main target of this view is to help users explain the generation of the event sequential patterns given the causal graph. To achieve this target, users can click on the button beside a sequential pattern (Fig.~\ref{system}(B2)) and a causation-aware novel design called \textbf{causal sequential flow} (Fig.~\ref{system}(B3)) will be shown to explain the sequential patterns. 

\textbf{Causal sequential flow.} This design integrates the information of sequential patterns as well as the causality of events inherently (G3). The sequential pattern (Fig.~\ref{flow_design}(A)) is placed horizontally on the top. Each event at the top has a corresponding circle below (Fig.~\ref{flow_design}(B)). The circles are connected as a line chart to show the chronological order of events and the vertical positions of circles encode the causality information between events. Specifically, an event's vertical position is lower than its parents' (i.e., the causes of this event) vertical position (Fig.~\ref{flow_design}(C)). To encode the information that event \textit{Block} is caused by \textit{Twist} and \textit{Flick}, the vertical region of \textit{Twist} and \textit{Flick} (Fig.~\ref{flow_design}(D)) will have a flow connecting to the region of \textit{Block}. Hence, from left to right, the design shows the temporal sequential patterns as a line chart. From top to bottom, the design encodes the causality information as a Sankey visualization. The length of the Sankey bar of events (Fig.~\ref{flow_design}(E)) is proportional to the number of causes. We do not show the Sankey bar for events that have no causes for the differentiation (Fig.~\ref{flow_design}(F)). With this design, users can clearly see the temporal information of events as well as the pathways that cause each event, which is useful for explaining the generation of event sequences.

\begin{figure}[htb]
	\centering
	\includegraphics[width=1\linewidth]{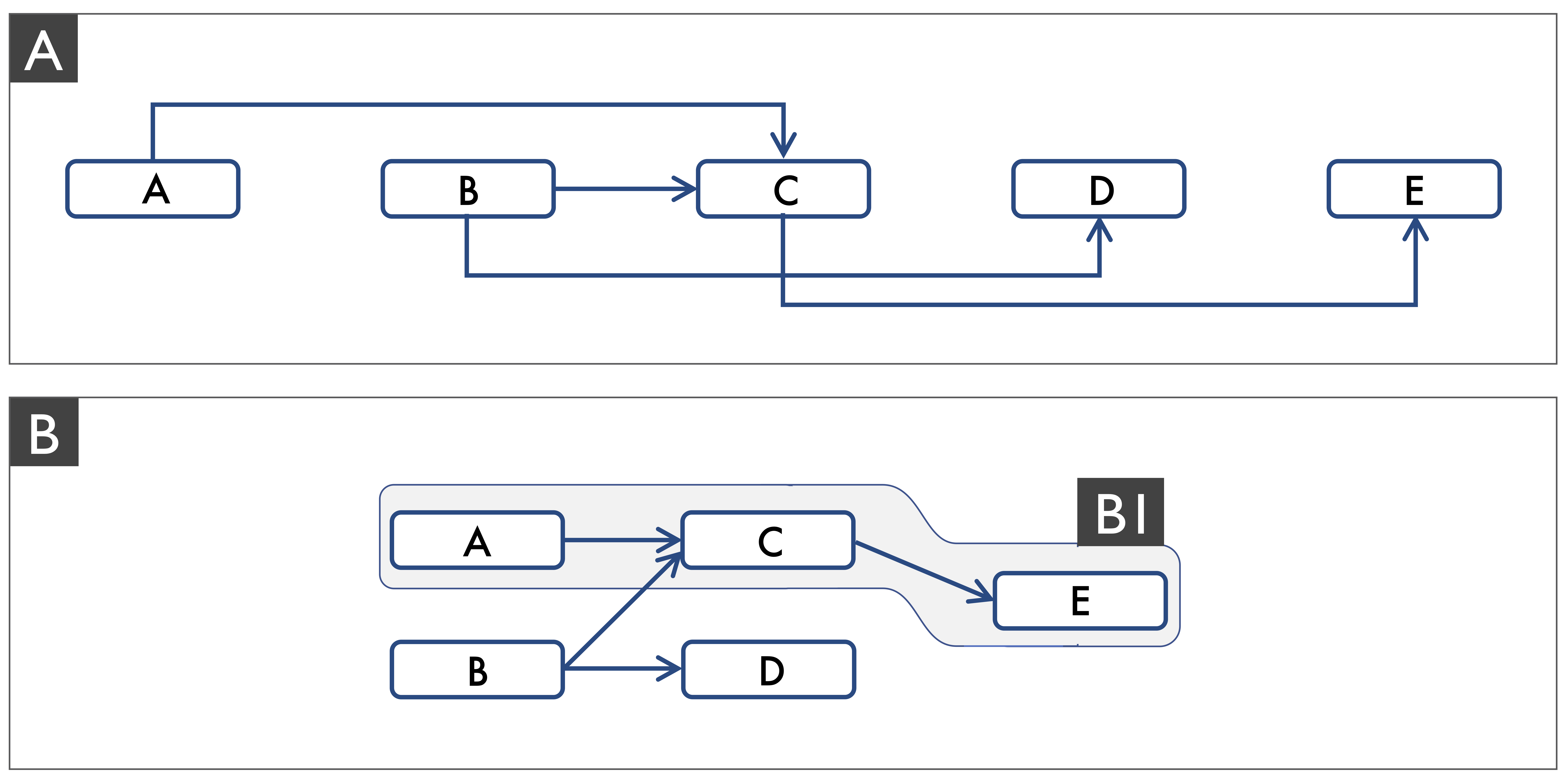}
    \caption{The design alternatives of causal sequential flow. (A) Encoding causal links in the sequential pattern visualization. (B) Encoding sequential patterns in the causal graph.}~\label{alternative}
    \vspace{-4mm}
\end{figure}

\textit{Design alternatives.} During the iteration of the visual design, we have discovered multiple design alternatives for explaining the sequential patterns with a causal graph. We prefer integrated views rather than separate views as it can provide more context for the understanding of causal information. Here we discuss the design of different integrated views. As shown in Fig.~\ref{alternative}(A), the first alternative is to directly draw causal links on the event sequential patterns. As this is intuitive when the causality information is simple, it will face an issue of crossing links when the causality information becomes complex. More importantly, with such visual representation, users cannot quickly identify the causal pathways. Users commented that although it is easy for them to find events that have links pointing to the targeted events, it is difficult to track the path of links which poses challenges for root cause analysis. The second alternative is to add the information of sequential patterns into the causal graph. As shown in Fig.~\ref{alternative}(B), in the causal graph, nodes of a sequential pattern will be covered by a context flow Fig.~\ref{alternative}(B1). This is inspired by Elzen and Wijk's \cite{sankeytree} design of decision trees. Although users commented that this is intuitive for the causality understanding, we find that this design is only suitable when the causality follows a tree structure. It is difficult to present the sequential pattern when the causal structure has multiple roots (e.g., pattern of $[A, B, C, ...]$ in Fig.~\ref{alternative}(A)). Hence, considering the usability and the comprehensibility, we decide to use the current design.

\begin{figure}[htb]
	\centering
	\includegraphics[width=1\linewidth]{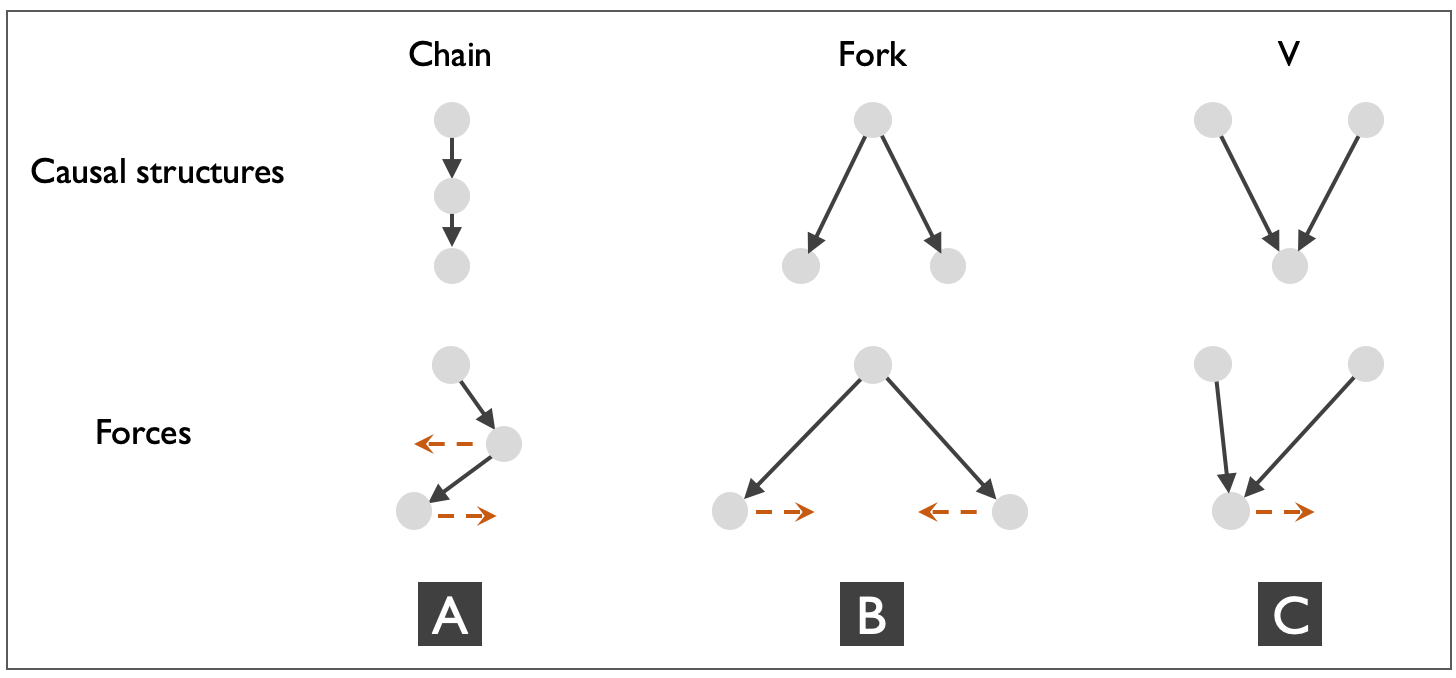}
    \caption{The layout algorithm of causal sequential flow. The causal graph mainly comprises of three causal structures, i.e., Chain (cascading causation), Fork (common causes), and V (common effects). A legible sequential flow is to ensure the readability of the these structures. Hence, we design three different forces to preserve these structures.}~\label{layout}
    \vspace{-4mm}
\end{figure}

\textit{Layout Algorithm.} Here we describe how to generate a legible causal sequential flow without severe visual clutter. The flow of Sankey could overlap with each other when the causality information is complex. To reduce the clutter, we need to determine the vertical and the horizontal position of the Sankey bar. For the vertical position, according to the encoding of sankey which has a consistent flow direction, we need to find an order of events $[e_1, e_2, ..., e_n]$ such that if $e_i$ is the cause of $e_j$, then $i < j$. Here we use Kahn's\cite{kahn1962topological} method that can derive such order of nodes for every DAG and set the default vertical position of events according to the order. Next, we need to set the horizontal position of each bar to show the causal relations. As shown in Fig.~\ref{layout}, we identify three causal structures according to literature \cite{tvcg/WangM16}, i.e. chain structure (Fig.~\ref{layout}(A)), fork structure (Fig.~\ref{layout}(B)), and v structure (Fig.~\ref{layout}(C)). 
A legible causal sequential flow is to ensure that users can easily perceive these structures. To this end, we use a force-directed approach to adjust the horizontal position of each Sankey bar. We propose three forces for three structures respectively (Fig.~\ref{layout}).
For the chain structure, we create forces for preventing the large slope (Fig.~\ref{layout}(A, bottom)) while for the fork and the v structure, we create forces for specifying appropriate angles (Fig.~\ref{layout}(B, C; bottom)).
Each Sankey bar will be applied to the three forces concurrently and we obtain the final horizontal position by an iterative process.
This layout process is flexible and it is easy to add additional layout criteria by introducing new forces.

\subsubsection{Data View}
When clicking a causal graph (Fig.~\ref{system}(A1)), the data view will be updated to show the sequences of the graph. Each row (Fig.~\ref{system}(C1)) shows an event sequence in which events are represented as circles and placed on the horizontal axis. Users can validate their hypothesis by inspecting the raw data. Users can further select a pattern in the sequential view (Fig.~\ref{system}(B1)) and the data view will show the sequences that match this pattern (Fig.~\ref{system}(C2)).

\subsubsection{Interaction}
Certain interactions are designed to support the analysis.

\textbf{Subgraph selection.} Users are allowed to select a subgraph in the causal graph view. This interaction is used when users are interested in only a set of events and would like to explore the causality within this set. By selecting a subgraph one of the causal graph in the causal graph view, the other event glyphs and links in all the causal graphs would be faded out. This can help users more easily compare the causality of a subset of events (G2). Concurrently, the sequential view would be updated according to the selected subgraph. The sequential patterns in the sequential view will be filtered and only the patterns that can be explained by the subgraph will be preserved. In addition to arbitrarily selecting a subgraph, users can also double click on an event glyph to select a subgraph that is composited by paths pointing to the targeted event for a target-based event analysis (G4).

\textbf{Sorting.} Users can sort the sequential patterns in the sequential view. By clicking the button (Fig.~\ref{system}(B3)), The sequential patterns will be sorted according to the statistics bar (Fig.~\ref{system}(B2)) in a descending order. This can help users focus on patterns with higher frequencies.

\section{Case Study} \label{sec:evaluation}
For the evaluation, we apply the system on two different dataset and invite experts to conduct two case studies respectively. 
Before the case studies, we introduce the system design to the experts and ask the experts to explore the system until they are familiar with the system usage.
After the case studies, we interview with the experts to collect their feedback about the system. The analysis process is done by the experts individually. We summarize the insights derived from the case studies and the experts' feedback as follows.

\begin{figure}[!htb]
	\centering
	\includegraphics[width=1\linewidth]{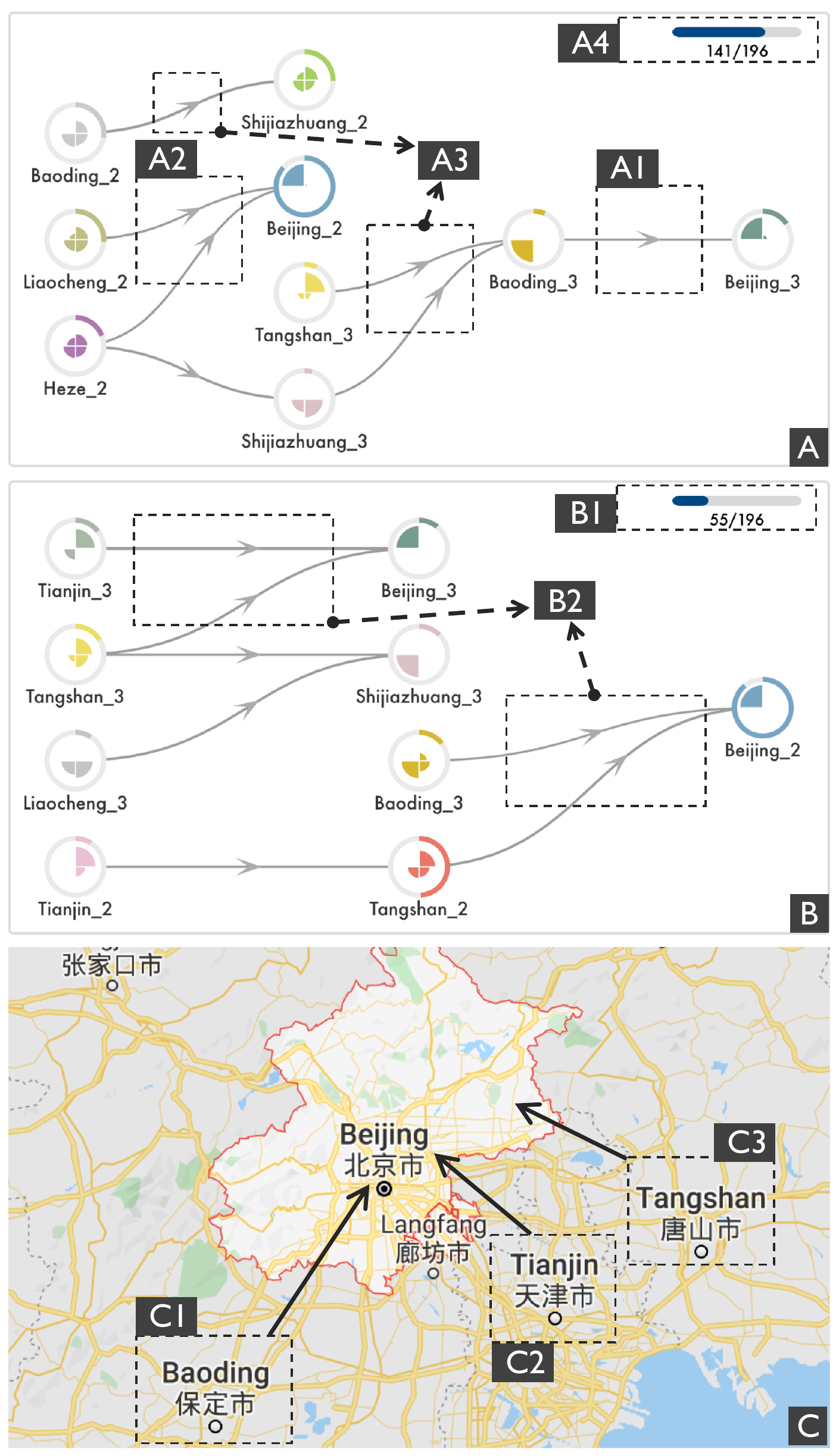}
    \caption{The causal graphs of two different air pollution propagation modes of Beijing and the explanations. (A) The causal graph which indicates that from a large set of Beijing's air pollution event sequences, Baoding contributes to the heavy haze of Beijing (A1) while Heze and Liaocheng contributes to the haze of Beijing (A2). (B) The causal graph of the rest of event sequences where Tangshan and Tianjin have contributions to the haze of Beijing (B2). (C) The geo explanations of the detected propagation patterns. Baoding (C1) is at the southwest while Tianjin (C2) and Tangshan (C3) is at the southeast of Beijing. The two causal graphs represent two different propagation patterns caused by different wind directions.}~\label{case1_1}
\end{figure}

\subsection{Air Pollution Logs}
In this case, we use the air pollution logs of Beijing and nearby cities (e.g., cities of Jing-Jin-Ji area) to analyze the source of Beijing's air pollution. 
We invited two experts to conduct analysis with the system. 
As an economic and political center of China, Beijing has suffered from heavy air pollution for a long time and the source of Beijing's air pollution is still a controversial topic.
To this end, our experts aim to obtain several insights by using our system.

The dataset record PM$_{2.5}$ of 11 cities every hour from January to June, 2016. 
According to prior studies of air pollution\cite{yan2015heavy}, we categorised the value of PM$_{2.5}$ into three levels, clean (0-75), haze (75-200), and heavy haze ($>$200). 
An event of air pollution is defined as a city's PM$_{2.5}$ increases to a higher level (\textit{to haze} and \textit{to heavy haze}) and we totally have 22 different events.
Since the target is to analyze the pollution source of Beijing, for each air pollution event of Beijing, we create an event sequence $[e_1, e_2, ..., e_n]$ by considering the preceding air pollution events of other cities.
A time session is set when including preceding events.
With this setting, we obtain 196 event sequences of Beijing's air pollution. The experts hope to characterize the relation between cities' air pollution events by analyzing these sequences.

At the start of the analysis, the experts skimmed over the causal graph view. Two causal graphs were presented and the experts first focused on the first causal graph with a longer statistics bar. In this graph, the experts looked for nodes of Beijing and found that for the heavy haze of Beijing, the heavy haze of Baoding is a direct cause (Fig.~\ref{case1_1}(A1)) while for the haze of Beijing, both haze of Liaocheng and Heze are the causes (Fig.~\ref{case1_1}(A2)). The experts commented that Baoding is a city of Hebei province while Heze and Liaocheng are from Shandong province. Baoding is comparatively more close to Beijing based on the geo distance. Moreover, with various heavy industries, the pollution from Baoding may contain a higher concentration of PM$_{2.5}$ and therefore caused the heavy haze of Beijing. This finding also matches with the pattern found by air pollution analysis\cite{yan2015heavy} in 2014, who indicated that cities of Hebei and Shandong had contributions to the air pollution of Beijing. The experts then explored the rest causal information in this graph. Most causal links (Fig.~\ref{case1_1}(A3)) indicate that there were air pollution transportation between cities of Hebei (Tangshan, Baoding, and Shijiazhuang), which is also consistent with the experts' domain knowledge. After investigating the graph, the experts turned to the sequential view by clicking the graph panel. From the sequential flow, the experts found that there were no long sequential patterns for the air pollution of Beijing. The experts hypothesized that the nearby cities of Beijing had severe air pollution created by themselves due to the large volumes of industries and it was hard to identify a clear propagation path that involves a set of cities.

The two statistic bars (Fig.~\ref{case1_1}(A4, B1)) indicated that the first causal graph was more commonly seen in the event sequence data. However, for a comprehensive analysis, the experts continued to investigate the second causal graph and compare it with the first graph. As indicated by the second graph (Fig.~\ref{case1_1}(B2, B3)), Beijing's air pollution (both haze and heavy haze) is related to Tangshan and Tianjin. The experts commented that Tangshan and Tianjin are in the southeast of Beijing (Fig.~\ref{case1_1}(C2, C3)) while Baoding is in the southwest of Beijing (Fig.~\ref{case1_1}(C1)). The two graphs may represent the pollution transportation of different wind directions. The experts stated that the division of causal groups and the causal relations seems reasonable. Further validation can be done by incorporating the meteorological data.

In this case, the experts obtain insights that the nearby cities of Beijing did have a contribution to the air pollution of Beijing while at most of the time, Baoding seemed to be an important source of Beijing's heavy air pollution.

\begin{figure}[!htb]
	\centering
	\includegraphics[width=1\linewidth]{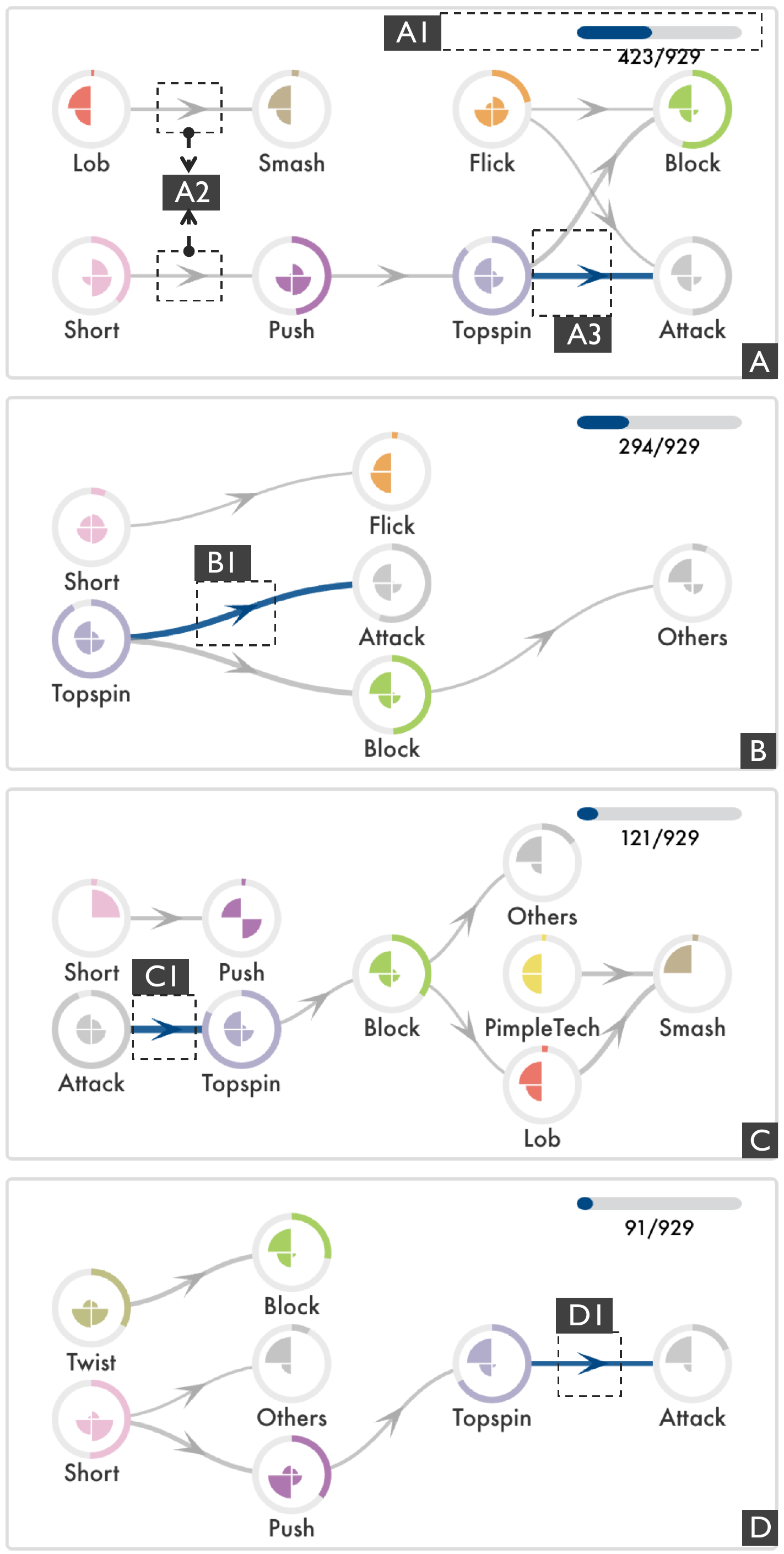}
		\caption{Four causal graphs (A, B, C, D) are detected from table tennis players' action sequences. (A) is the causal graph that cover the largest set of sequence data (A1). Causal relations that already know (A2) and new insights about the relation of techniques (A3) are identified. When comparing between the graphs, it is found that (B) and (D) share the same link (Topspin to Attack, blue) with (A) while in (C), this link has a reversed direction (C1). }~\label{case2_1}
		\vspace{-4mm}
\end{figure}

\subsection{Action Sequences of Table Tennis }
In this case, we use this system to investigate the causality relation between table tennis players' action sequences. The dataset we used contains 9 matches, which are all Ito Mima against Chinese players. The event sequences that we analyzed is the stroke sequences of players. Each stroke has a set of attributes, including stroke placement, stroke position, stroke technique, and the stroke player etc. We invited two experts (referred as A and B) to conduct the analysis. Expert A is a professor of sports science and has long time experiences of the tactical analysis of table tennis. Expert B is a Ph.D. candidate who majors in the sports analytics of table tennis. According to the experts' interests, we transform the stroke sequences to sequences of stroke techniques $[e_1, e_2, ..., e_n]$ where $e_i$ is the stroke technique of the $i$th stroke. There are 16 different stroke techniques. With the system, the experts are eager to identify the causal relationship between the adoption of different stroke techniques and answer causal questions such as why would a player use certain stroke techniques.

The experts started with the causal graph view and noticed four causal groups. They first sorted the causal groups according to the number of sequence instances. From the statistics bar (Fig.~\ref{case2_1}(A1)), they realized that most of the action sequences belong to the first causal group. By inspecting the graph of this group, they identified a set of causal relations that they already known, such as \textit{Lob} to \textit{Smash}, \textit{Short} to \textit{Push}, etc (Fig.~\ref{case2_1}(A2)). The experts commented that this strengthened their confidence in the causal detection result. They further investigate the rest causal relations in the graph.  

With the donut of event glyph (Fig.~\ref{case2_1}(A)), they noticed that \textit{Push}, \textit{Topspin}, \textit{Block} and \textit{Attack} are the most frequently used stroke techniques. Specifically, the links between \textit{Topspin}, \textit{Block} and \textit{Attack} (Fig.~\ref{case2_1}(A3)) showed that the use of \textit{Topspin} is a common cause of the use of \textit{Attack} and \textit{Block}. The experts commented that these three techniques are frequently used when two players are in a stalemate. \textit{Topspin} and \textit{Attack} are offensive techniques while \textit{Block} is a defensive technique. Hence, the link between \textit{Topspin} and \textit{Block} is expected, as it represents that one player is trying to defend the attack from another player. 
However, for the link between \textit{Topspin} and \textit{Attack}, they previously knew that the two techniques are highly correlated but not clear about the direct causal relation between them. The finding of the causal graph provided them a clear relation. Moreover, the experts explained that Ito is not good at using \textit{Topspin} and would prefer to use \textit{Attack} when facing a stalemate. Hence, the causal direction of the two techniques indicated that Chinese players' usages of \textit{Topspin} forced Ito to use \textit{Attack}, showing that Chinese players controlled the game situation. This is consistent with the game result of the matches (Chinese players won 7 of 9).

\begin{figure}[!htb]
	\centering
	\includegraphics[width=1\linewidth]{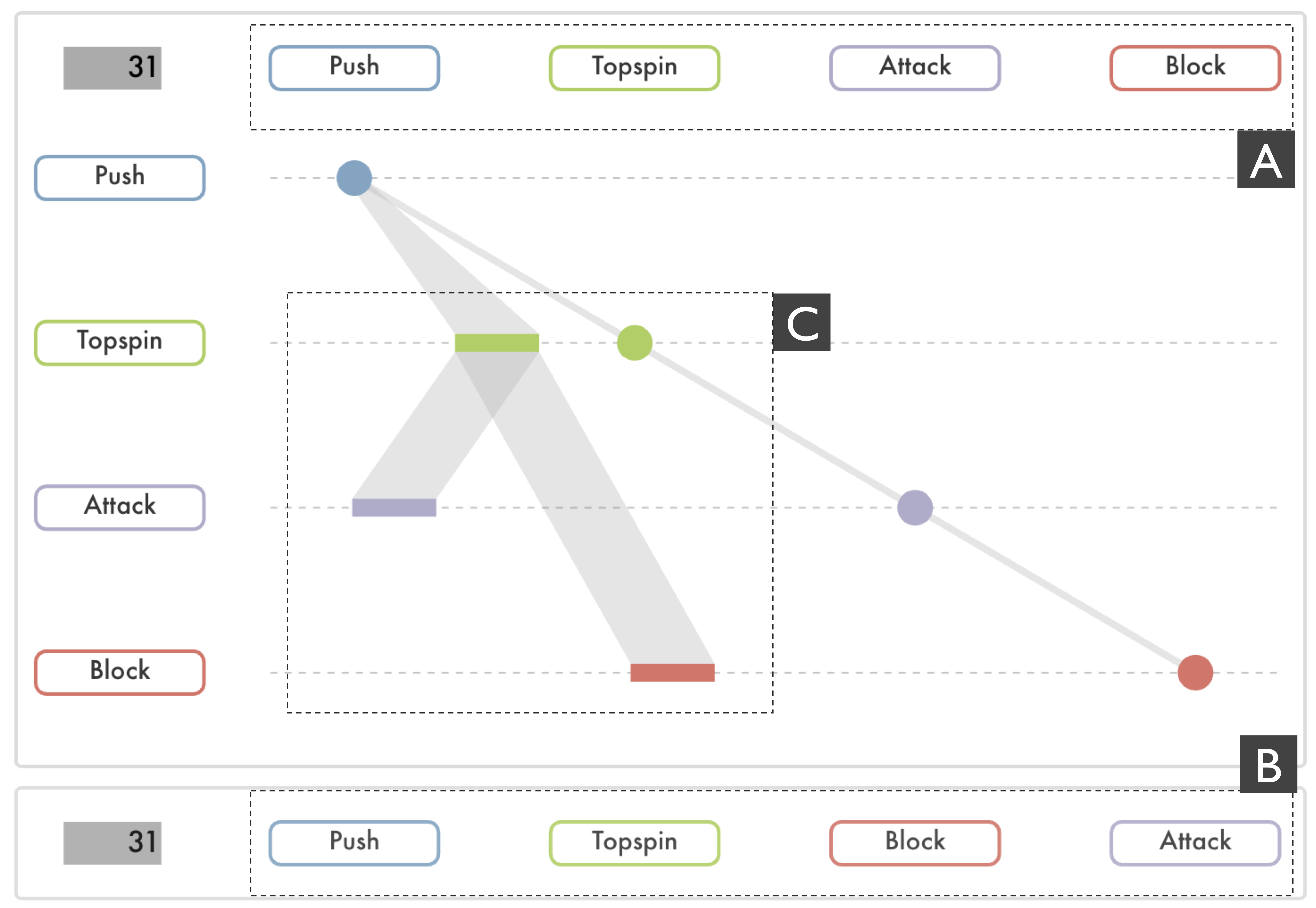}
		\caption{A and B represent two contradictory sequential patterns of a causal graph. The sequential flow (C) shows that this contradictory is created since Topspin is a common cause of Block and Attack. }~\label{case2_2}
		\vspace{-4mm}
\end{figure}

After exploring the first causal graph, the experts continued to compare the four causal graphs. The experts were interested with the discovered causal relation between \textit{Topspin} and \textit{Attack}. Hence, they hovered on this link (Fig.~\ref{case2_1}(A3)) for further investigation. From the highlighted links, the experts identified two identical links (Fig.~\ref{case2_1}(B1), (D1)) and a reversed link (Fig.~\ref{case2_1}(C1)) from \textit{Attack} to \textit{Topspin}.  According to the experts' hypothesis, this graph (Fig.~\ref{case2_1}(C)) may represent the rallies that Ito controlled the game situation, which is different from the other graphs.  Based on these graphs and the correspodning statistic bars, the experts hypothesized that Chinese players had dominated the game by recognizing Ito' weak point of dealing with \textit{Topspin}.

The experts then tried to use the causal graph to explain the observed sequential patterns of stroke technique sequences. Since the experts are more interested in the dominant rallies of Chinese players, they selected graph 1 (Fig.~\ref{case2_1}(A)) and jumped to the sequential view. In the sequential view, they noticed four different patterns. Among these patterns, two patterns seemed to be contradictory (Fig.~\ref{case2_2}(A, B)), as the order of \textit{Block} and \textit{Attack} was different. By inspecting the causal flow of the two patterns (Fig.~\ref{case2_2}(C)), the experts realized that the two different patterns are created since both \textit{Block} and \textit{Attack} were the result of \textit{Topspin}. The experts hypothesized that Ito was trying to deal with \textit{Topspin} with \textit{Block} and \textit{Attack} while the use of \textit{Block} and \textit{Attack} was based on the circumstances and had no determined orders. They further turned to the data view and verified their hypothesis.

In this case, the experts obtained the insights that Chinese players' skillful \textit{Topspin} is an important factor for the win of the games.

\section{Discussion} \label{sec:discussion}

\textbf{Significance.} While answering causal questions has been regarded as an essential task of data analysis, analysts have been struggling to conduct causation analysis with observational data without doing controlled experiments. Current practices of causal analysis majorly focus on the investigation of multi-dimensional data while few have studied the causation analysis of event sequences data. In this work, we introduce how to integrate state-of-the-art automatic causal discovery techniques into event sequence analysis and design novel sequence visualizations for conducting the causal analysis. The case studies have demonstrated that applying causal analysis to event sequence data can help users focus on a set of essential causal relations of events rather than a number of co-occurrence relations.

\textbf{Scalability.} We discuss the scalability of the length of event sequences, the number of causal groups, and the number of unique events. The design can support the exploration of long event sequences as we visualize the frequent sequential patterns (whose lengths are much shorter than the raw event sequences) in the sequential view. For the number of causal groups, the scalability is limited since we use juxtaposition in the causal graph view for the visual comparison. One solution to improve the scalability is to apply hierarchical clustering to the dataset so that users can first explore the high-level causal groups and decompose certain high-level groups to low-level subgroups on demand. The number of unique events is also limited due to the layout of the causal graph. According to our experiments, the layout can support the presentation of 20 - 30 events without much visual clutter. However, for those dataset with hundreds of different events, a larger visual space is required for the causal graph view to clearly present the causal relations. We have identified two different directions to improve the scalability. One is to group numerous events into a small set of categories. For example, we can group sports players' actions into categories such as defense and offense and identify the causal relation between the event categories. The other is to extend current large graph visualizations for a scalable causal graph layout.

\textbf{Limitations.} In this work we identify two major limitations. The first limitation is the lack of including descriptive information into the causal analysis of event sequences data. Take user behavior log analysis as an example. In addition to the behavior log itself, analysts are also interested in analyzing the causal impact of users' personal information, such as age, income, and gender etc., on their behaviors. This can help users more easily characterize the causal groups and obtain more clear insights. However, the proposed causal detection framework only considers the relation between events. In the future, we plan to integrate the descriptive information into the initial clustering step and estimate the causal impact of these information on the generation of events by using intervention techniques\cite{ijon/Shanmugam01}. The second limitation is about the explanation of the detected causal relation. Currently, as shown in our cases, the explanation is largely depended on our experts' domain knowledge and experiences. This is due to the lack of data dimensions. For example, for the air pollution case, by integrating the wind direction into the causal detection process can significantly facilitate the causal explanation and validation. We plan to collect more data and conduct more case studies in the future.
\section{Conclusion}

This study proposes an interactive visual analytics approach for analyzing the causation of event sequences. This is the first attempt to integrate automatic causal discovery techniques into the causation analysis of event sequences. We propose a causal discovery framework for identifying the causal relation of events and create a visual design named causalflow to inherently present the causal information and the chronological information of event sequences. We further implement a visual analytics system to help users comprehensively explore and anlayze the causation of events in a multi-level manner.

In the future, we plan to improve this work by enhancing the evaluation process of the event causal discovery and the scalability of the approach. For the evaluation, we plan to apply the model on more dataset and invite experts to validate the detected causal relations. For the scalability, we plan to integrate the hierarchical clustering technique into the causal discovery for handling event sequence dataset with a large set of events and incorporate theories of large graph visualizations and the visual summarization of event sequences to increase the visual scalability.

\end{spacing}


\bibliographystyle{abbrv-doi}

\bibliography{main}
\end{document}